\documentclass{article}

\usepackage{arxiv}

\usepackage[utf8]{inputenc} 
\usepackage[T1]{fontenc}    
\usepackage{hyperref}       
\usepackage{url}            
\usepackage{booktabs}       
\usepackage{amsfonts}       
\usepackage{nicefrac}       
\usepackage{microtype}      
\usepackage{lipsum}
\usepackage{graphicx}
\graphicspath{ {./images/} }

\usepackage{comment}
\usepackage{tabularx} 
\usepackage{float}
\usepackage{subcaption} 

\title{A Graph Theoretic Approach for Exploring the Relationship between EV Adoption and Charging Infrastructure Growth}

\author{
 Fahad S. Alrasheedi \\
  Department of Computer Science\\
  University of Nebraska at Omaha\\
  Omaha, NE 68182 \\
  \texttt{falrasheedi@unomaha.edu} \\
   \And
 Hesham H. Ali \\
  Department of Computer Science\\
  University of Nebraska at Omaha\\
  Omaha, NE 68182 \\
  \texttt{hali@unomaha.edu} \\
  }

\begin{document}
\maketitle
\begin{abstract}
The increasing global demand for conventional energy has led to significant challenges, particularly due to rising CO$_2$ emissions and the depletion of natural resources. In the U.S., light-duty vehicles contribute significantly to transportation sector emissions, prompting a global shift toward electrified vehicles (EVs). Among the challenges that thwart the widespread adoption of EVs is the insufficient charging infrastructure (CI). This study focuses on exploring the complex relationship between EV adoption and CI growth. Employing a graph theoretic approach, we propose a graph model to analyze correlations between EV adoption and CI growth across 137 counties in six  states. We examine how different time granularities impact these correlations in two distinct scenarios: \textit{Early Adoption} and \textit{Late Adoption}. Further, we conduct causality tests to assess the directional relationship between EV adoption and CI growth in both scenarios. Our main findings reveal that analysis using lower levels of time granularity result in more homogeneous clusters, with notable differences between clusters in EV adoption and those in CI growth. Additionally, we identify causal relationships between EV adoption and CI growth in 137 counties, and show that causality is observed more frequently in \textit{Early Adoption} scenarios than in \textit{Late Adoption} ones. However, the causal effects in \textit{Early Adoption} are slower than those in \textit{Late Adoption} 
\end{abstract}


\section{Introduction}


As global populations grow, the demand for conventional energy continues to rise, leading to increased CO$_2$ emissions and depletion of finite natural resources \cite{atabani2011review}. In the U.S., transportation contributes nearly 29\% of carbon emissions, with light-duty vehicles responsible for 59\% of these. This highlights the urgency of transitioning to sustainable energy sources \cite{Us_Epa1, Us_Epa2}.


Achieving zero carbon emissions in light-duty transportation is a crucial step toward this goal, and electric vehicles (EVs) offer a promising solution, as they produce zero tailpipe emissions \cite{center2022electric}. However, widespread EV adoption faces several challenges, including insufficient charging infrastructure, high upfront costs, and concerns about limited driving range \cite{chen2023optimal}. To mitigate these barriers, government interventions have been implemented, such as subsidies for charging infrastructure (CI) and tax credits for EV purchases. Therefore, a sufficient charging infrastructure is significant to sustain an effective transition to EV \cite{usa5}.

Moreover, several studies have investigated the influence of charging infrastructure alongside other factors, such as socio-demographic, economic, environmental, and political variables, on EV adoption in the United States \cite{usa1,usa2,usa3,usa4}. While charging infrastructure consistently emerges as a significant driver of EV adoption, the sole relationship between EV adoption and charging infrastructure growth remains underexplored. Existing research typically embeds charging infrastructure within broader multivariable models, making it difficult to isolate its unique effect. 

Thus, a key question remains open: Does widespread EV adoption stimulate the development of more charging infrastructure (\textit{Early Adoption}), or does the availability of charging infrastructure drive EV adoption (\textit{Late Adoption})? Additionally, such relationships may vary depending on the temporal granularity considered. Hence, this paper contributes to the field by investigating the sole relationship between EV adoption rate and CI growth in two different scenarios of \textit{Early Adoption} and \textit{Late Adoption} (as defined in Section \ref{cases}) through two complementary analyzes: (1) correlation analysis using a graph-theoretic approach and (2) causality analysis. 

In the correlation analysis, we propose a graph model for analyzing EV adoption patterns across 137 counties in six U.S. states, where nodes represent counties and edges represent correlations between their EV adoption rates. In each adoption scenario, we construct graph models at multiple time granularities, monthly, bi-monthly, quarterly, and bi-annual to capture temporal dynamics. Using the Louvain method, a community detection technique, we cluster counties into groups based on their correlation structures. A similar graph-theoretic analysis is performed on CI growth patterns. Subsequently, we compute the correlation between the EV adoption and CI growth at different combinations of adoption scenarios, and time granularities. In the causality analysis, we examine the directional causality relationship between EV adoption and CI growth for each county incorporating different time lags, and using the same combinations of adoption scenarios, and time granularities. 


Firstly, the study analysis reveals that, within EV adoption networks, counties from the same state tend to cluster together, exhibiting greater homogeneity at finer time granularities. However, this pattern is not observed in CI growth networks. Secondly, the correlations between EV and CI networks in both adoption scenarios, \textit{Early Adoption} and \textit{Late Adoption}, are consistently weakly positive, with \textit{Late Adoption} correlations being slightly higher than those in \textit{Early Adoption}. Thirdly, 115 out of 137 counties exhibit causal relationships between EV adoption and CI growth, analyzed using various combinations of time granularities, time lags, and the two adoption scenarios. Notably, causality is observed more frequently in \textit{Early Adoption} scenarios than in \textit{Late Adoption} ones. However, the causal effects in \textit{Early Adoption} are slower than those in \textit{Late Adoption}, which can be attributed to the significantly higher costs of establishing new CIs compared to adopting new EVs.

The remainder of this paper is organized as follows. In Section \ref{lit}, we review the related work to the EV adoptions and CI growth. Section \ref{metho} discusses our approach for employing a graph-theoretic approach to build the EV adoption’s correlation networks, and the CI growth's correlation networks, and for the causality tests. Finally, Section \ref{result} discusses the results followed by the conclusion and future work in Section \ref{conclusion}.

\section{Literature Review}
\label{lit}

In this section, we review existing research on the complex relationship between EV adoption and CI growth. In addition, we discuss studies that have utilized a graph model in the context of CI optimization and identify gaps in the literature that our study aims to address. This review sets the stage for understanding the unique contributions of our research to the field.

\subsection{Impact of Charging Infrastructure and Other Factors on EV Adoption}
\label{CI_lit}

There is an extensive body of literature examining electric vehicle (EV) adoption globally. Since our study focuses on the United States, we narrow our review to studies that are most relevant to our analysis—specifically, those conducted within the U.S. that consider charging infrastructure as a key factor influencing EV adoption. These studies provide the necessary background and support for our research approach. To underscore the broader significance of this topic, we also briefly highlight selected national-level studies from other countries at the end of this section.


Ledna \textit{et al.} investigated the impact of various support policies on EV adoption in California, focusing on investments in public charging infrastructure and vehicle purchase subsidies \cite{ledna2022support}. They simulated three policy scenarios: no support, support for either infrastructure or vehicle purchases, and combined support; under both conservative and optimistic assumptions regarding technological advancement. Their findings suggest that combined support policies are most effective in boosting EV sales and reducing CO$_2$ emissions overall, while infrastructure-focused support yields better outcomes in conservative technology scenarios, emphasizing the importance of tailoring policy strategies to market and technological conditions.


In another study, Chen \textit{et al.} developed a game-theoretic model to examine the strategic interactions among governments, technology firms, and consumers in promoting electric vehicle (EV) adoption under various subsidy policies \cite{chen2023optimal}. Their goal was to identify optimal government incentive strategies, focusing on infrastructure investment subsidies versus usage subsidies, under constrained budgets. The study found that when the budget is limited, either type of subsidy can be effective. However, under a high-budget scenario, investment subsidies are more effective for maximizing market penetration, while usage subsidies better enhance consumer surplus and overall social welfare


White \textit{et al.} examined the mechanisms underlying the relationship between public charging infrastructure and electric vehicle (EV) adoption intent across three major U.S. metropolitan areas: Los Angeles, Dallas/Fort Worth, and Atlanta \cite{white2022charging}. Using multiple mediation analysis, they tested whether three psychological constructs—range anxiety, perceived mobility restriction (PMR), and subjective norms—mediated the effect of public charging station density on EV adoption intent. Their findings revealed that subjective norms were the strongest and most consistent mediator, significantly influencing adoption intent in all three regions. Range anxiety played a marginal role in two of the regions, while PMR showed no significant mediating effect. These results suggest that the visibility of charging infrastructure may function more as a social signal promoting adoption, rather than alleviating technical or mobility concerns.


Khan \textit{et al.} examined the distribution of EV charging infrastructure across New York City to identify the socio-demographic and transportation factors associated with charging station access \cite{khan2022inequitable}. Using correlation and hypothesis testing at the zip-code level, they found that the presence of highways, higher median household income, and a greater percentage of White-identifying population were positively associated with the presence and number of charging stations. Their findings highlight significant disparities, with low-income and minority communities having notably less access to EV charging infrastructure, underscoring the need for equity-focused policies in infrastructure planning.


Burra \textit{et al.} addressed the lack of sufficient EV-owning households in Maryland’s travel surveys by developing a synthetic population using a Bayesian network approach \cite{usa1}. They estimated household-level EV ownership probabilities using a binomial logit model and evaluated how socio-demographic factors and access to Level 2 and DC fast charging infrastructure influence adoption. Their findings revealed that high-income and suburban households are more likely to own EVs, and that access to workplace charging and DC fast chargers significantly increases the likelihood of ownership. The study highlights the importance of differentiating between charger types and suggests that equitable infrastructure deployment should consider both location and income disparities.


Pamidimukkala \textit{et al.} conducted a survey-based study of University of Texas at Arlington affiliates to examine how four categories of barriers—technological, environmental, financial, and infrastructure—affect consumers’ intentions to adopt electric vehicles (EVs) \cite{usa2}. Using structural equation modeling on responses from 733 participants, they found that financial, infrastructure, and technological barriers all had significant negative effects on EV adoption intention, with financial barriers having the strongest influence. Infrastructure barriers—such as insufficient public charging stations and limited maintenance and repair services—also emerged as key obstacles. In contrast, environmental barriers were not statistically significant, suggesting that respondents largely accepted the environmental benefits of EVs.


Debnath \textit{et al.} applied computational text analysis to approximately 36,000 Facebook posts—comprising a corpus of around 600,000 words—to explore public discourse surrounding electric vehicle (EV) adoption in the United States across six PESTLE dimensions: Political, Economic, Social, Technological, Legal, and Environmental \cite{usa6}. Using a mixed-methods approach that combined social network analysis and machine learning-based topic modeling, they identified key themes within each category. The analysis revealed that discussions of charging technology and renewable energy dominated the Technological dimension, while concerns related to charging infrastructure—particularly public funding, regulation, and policy—were primarily clustered under the Political category.

Kamis \textit{et al.} developed predictive models for three 2021 targets—EV registrations, EV-related jobs, and new public charging infrastructure (Level 2)—using county-level data across a wide range of predictors, including existing charging infrastructure, environmental conditions, education, and socio-demographic factors. They found that existing charging infrastructure was a significant predictor across all models, with Gradient Boosted Trees showing superior accuracy over traditional regression methods \cite{usa3}.


At the national level, Broadbent \textit{et al.} developed a national-scale policy simulation model to explore how public incentives and infrastructure investment affect electric vehicle (EV) adoption trajectories and emissions reductions in Australia. Additionally, Babic \textit{et al.} proposed a simulation-based operational model for transforming urban parking lots in Melbourne into EV-enabled charging hubs, focusing on profitability, charger allocation, and consumer behavior \cite{babic2022data}. In Germany, Illmann \textit{et al.} analyzed monthly ZIP-code-level data from 2012 to 2017 and found a significant long-term relationship between public charging infrastructure and private EV registrations, with fast chargers having the strongest influence \cite{illmann2020public}. In France, Haidar \textit{et al.} used mixed-effects regression to analyze BEV and PHEV adoption across 94 departments and found that different incentives and charging infrastructure types affect each market differently, with fast and ultra-fast chargers driving BEV sales, and slow-to-normal chargers more relevant for PHEVs \cite{haidar2022relationship}. In China, studies examine the relationship between electric vehicle (EV) adoption and the availability of charging infrastructure. They consistently find that improvements in charging infrastructure have a positive impact on promoting EV adoption, though the nature of the effect depends on the policy design, regional conditions, and stakeholder interactions \cite{usa7,usa8}.

\subsection{Application of Graph Model to Charging Infrastructure}
\label{GH_lit}

The graph model is a powerful tool for analyzing relationships between various entities. A correlation network, where nodes represent entities or variables, and edges indicate correlation strength, is widely used in diverse fields, including bioinformatics, engineering, and social sciences \cite{corr1,corr2,corr3}.


The application of graph-theoretic models to electric vehicle (EV) charging infrastructure has predominantly focused on optimizing station placement and network coverage. Gagarin \textit{et al.} modeled the problem as a multiple domination problem on reachability graphs to ensure drivers can access multiple charging options within a limited cruising range \cite{gagarin2018multiple}. Arkin \textit{et al.} employed unit disk graphs and t-spanners to determine the minimum number of stations required to enable near-shortest EV routes with sufficient recharging support \cite{arkin2019locating}. Moreover, studies in \cite{funke2015placement,funke2016placement, agarwal2016efficient} applied the concept of the hitting set to identify the optimal number and locations for charging stations. Finally, Altundogan \textit{et al.} integrated graph-based distance modeling with a genetic algorithm to optimize the spatial distribution of a fixed number of chargers across urban networks \cite{altundogan2021genetic}. 

While these studies demonstrate the utility of graph theory for infrastructure planning, they do not explore the dynamic, temporal interplay between EV adoption trends and charging infrastructure growth—an area this study addresses by applying graph models to correlate and interpret adoption patterns directly.

\subsection{Gaps Found in the Literature}
\label{gaps}

As we discussed the literature in sections \ref{CI_lit} and \ref{GH_lit}, we noticed two gaps in the literature which the current study addresses:

\begin{enumerate}
\item Most of the existing literature examines EV adoption and its relationship with other factors in one to three cities, resulting in limited data diversity. This study, however, analyzes 137 counties across six U.S. states, enhancing the analysis and demonstrating the role of states in clustering counties within the correlation networks.

\item The use of graph model has mostly been limited to optimizing the placement of charging stations within specific areas. In contrast, this study contributes by applying graph model to explore patterns in both EV adoption and CI growth networks under two scenarios of \textit{Early Adoption} and \textit{Late Adoption} , build a correlation network, and cluster counties accordingly.


\item Our analysis examines two scenarios: \textit{Early Adoption} and \textit{Late Adoption}. We use different time granularities and lags to capture complex correlations between counties in both EV adoption and CI growth, as well as to identify potential causal relationships between EV adoption and CI growth within each county.

\end{enumerate}

\section{Methodology}
\label{metho}

In this section, we outline the approach used to analyze the relationship between EV adoption and charging infrastructure (CI) growth. It starts by describing the data in Section \ref{Data}, using graph model to build correlation networks, introducing the two cases of adoption in Section \ref{cases}, setting different ime granularities in Section \ref{seasons}, and setting two causality tests in Section \ref{cause}

\subsection{Data Collection}
\label{Data}
\subsubsection{Electric Vehicles (EV) Data}

\begin{table}[]
\centering
\caption{Number of Completed Counties by State}
\begin{tabular}{|l|l|l|}
\hline
No & State & \# of Completed Counties \\ \hline
1 & Colorado & 20 \\ \hline
2 & Minnesota & 3 \\ \hline
3 & Montana & 2 \\ \hline
4 & New York & 48 \\ \hline
6 & Texas & 30 \\ \hline
7 & Virginia & 34 \\ \hline
8 & Total & 137 \\ \hline
\end{tabular}
\label{table: table00}
\end{table}

The data used for electric vehicle (EV) registrations was obtained from Atlas Hub\cite{Nigro_undatedj}; which  works directly with states to provide comprehensive temporal data across various U.S. states, primarily at the zip code and county levels. For this study, we selected states offering consistent data from 2018 to 2023 at the monthly basis; so counties that had 12 months of data for each year in the selected period were included in the final dataset. Consequently, this dataset consists of EV data from 137 counties across six states; we aggregate the data at various time granularities for analyzing the data at different time scales. Table \ref{table: table00} provides a summary of the states and the number of counties from each. 


\subsubsection{Charging Infrastructure (CI) Data}

Data on public and private charging stations were collected from the Alternative Fueling Station Locator, which, at the time of this research, included a total of 67,540 stations \cite{noauthor_undated}. To ensure consistency, the dataset was filtered to include only the counties represented in the EV data. More importantly, the CI for a county is represented by the number of chargers rather than the number of charging stations, as it more accurately reflects the infrastructure's capacity. Consequently, our analysis investigates the relationships between EV adoption and different CI levels, including Level 2 chargers, Direct Current (DC) chargers, and a combined level that includes Level 1, Level 2, and DC chargers, referred to as “ALL.” Additionally, each CI level for a county is aggregated into various time granularities, consistent with the approach used for the EV adoption data.

\subsection{Proposed Graph Theoretic approach: correlation network }
\label{corrNet}



To investigate the structural relationships among counties based on their electric vehicle (EV) adoption or charging infrastructure growth, we construct an \textbf{unweighted, undirected graph} where nodes represent counties and edges represent statistically significant correlations between them. Let $C_i$ and $C_j$ denote the time series associated with counties $i$ and $j$, respectively. The similarity between each pair of counties is measured using the Pearson correlation coefficient:
\[
\rho_{ij} = \frac{\mathrm{Cov}(C_i, C_j)}{\sigma_{C_i} \sigma_{C_j}},
\]
where $\mathrm{Cov}(C_i, C_j)$ is the covariance between the two time series, and $\sigma_{C_i}$ and $\sigma_{C_j}$ are their standard deviations. We construct a graph $G = (V, E)$, where each node $v \in V$ represents a county, and an edge $(i, j) \in E$ exists if $\rho_{ij} \geq \tau$, with $\tau$ being a correlation threshold selected through a modularity optimization process.

To identify community structure within the network, we apply the \textit{Louvain community detection algorithm}, which partitions the graph into non-overlapping communities. The quality of the partition is assessed using the modularity metric, defined as:
\[
Q = \frac{1}{2m} \sum_{i,j} \left[ A_{ij} - \frac{k_i k_j}{2m} \right] \delta(c_i, c_j),
\]
where $A_{ij}$ is the adjacency matrix of the graph, $k_i$ and $k_j$ are the degrees of nodes $i$ and $j$, $m$ is the total number of edges, $c_i$ is the community assignment of node $i$, and $\delta(c_i, c_j)$ is the Kronecker delta function, equal to 1 if $c_i = c_j$ and 0 otherwise~\cite{newman2004finding}. The correlation threshold $\tau$ is selected to maximize the modularity $Q$ and ensure meaningful community separation.

After constructing the initial graph and performing community detection, we apply a \textit{refinement procedure} that iteratively removes the edge with the highest \textit{edge betweenness centrality}, defined as the number of shortest paths passing through that edge. After each removal, the modularity is recalculated to evaluate improvements in community structure. This process helps expose more distinct and internally cohesive communities by removing structural bridges between densely connected groups of counties.

\subsection{Early versus Late Adoption}
\label{cases}

We define two adoptions scenarios of \textit{Early Adoption} and \textit{Late Adoption} to examine the relationship between EV adoption and CI growth using different time granularities.

\subsubsection{Early Adoption}
In this scenario, we assume that EVs are adopted before sufficient CI is available. In other words, EV adopters take the risk of using EVs despite a lack of supporting infrastructure, expecting their adoption to prompt further CI development. To analyze this, we apply a one-year lag, where EV adoption data spans 2018 to 2022 while CI growth data covers 2019 to 2023. This means a given year in the EV data (e.g., 2020) is aligned with the following year in the CI data (e.g., 2021). In this scenario, we investigate whether correlations exist between EV adoption and each level of CI, accounting for the one-year lag between the two datasets.

\subsubsection{Late Adoption} 
This scenario assumes that CI is established before widespread EV adoption, with the expectation that sufficient CI will encourage more consumers to adopt EVs. Similar to the \textit{Early Adoption} scenario, we apply a one-year lag: CI growth data spans 2018 to 2022, while EV adoption data covers 2019 to 2023. Here, CI data for a given year is aligned with EV data from the following year (e.g., 2018 CI data is aligned with 2019 EV data.)

\subsection{Different Time Granularities}
\label{seasons}



The relationship between electric vehicle (EV) adoption and charging infrastructure (CI) growth is complex; therefore, our analysis explores multiple time granularities to capture underlying patterns that may emerge at different temporal resolutions. To this end, we divide the dataset into several time granularities: monthly, bi-monthly, quarterly, and bi-annual periods. For each granularity, we calculate the growth rate from one period to the next. For instance, at the monthly level, we compute the adoption or growth rate from January to February, February to March, and so on. At the quarterly level, we calculate growth from the first quarter to the second quarter, and so forth. The adoption or growth rate (AGR) at each step is mathematically expressed as:

\[
\mathrm{AGR}_{g_j} = \frac{g_j - \sum_{k=1}^{j-1} g_k}{\sum_{k=1}^{j-1} g_k}
\]

where \( \mathrm{AGR}_{g_j} \) represents the growth rate at granularity \( g_j \), \( g_j \) is the number of registered EVs or newly established CI units during the current period, and \( \sum_{k=1}^{j-1} g_k \) is the accumulated total from all previous periods.

Through this analysis, we examine the correlation networks of EV adoption and the growth of \textbf{multiple CI levels} (e.g., Level 1, Level 2, and DC fast charging) across different temporal granularities. Specifically, we investigate whether clusters identified at finer resolutions persist when observed at coarser levels, or whether new inter-county correlations emerge, thereby shedding light on the temporal dynamics of EV adoption and CI expansion across different types of infrastructure.

\subsection{Causality Relationships between EVs and CIs}
\label{cause} 

To explore the causal relationship between EV adoption and CI growth, we formulate two hypotheses:

\begin{itemize}
\item \textit{Early Adoption} Scenario: In this case, we test whether EV adoption leads to the development of CI, expressed as Null Hypothesis 1: "EV adoption does not cause CI growth." We evaluate this hypothesis using the Granger causality test.

\item \textit{Late Adoption} Scenario: Here, we test whether CI growth drives EV adoption, expressed as Null Hypothesis 2: "CI growth does not cause EV adoption." We evaluate this hypothesis using the Granger causality test.
\end{itemize}

The Granger causality test is applied with an alpha level of 0.05, testing whether past values of one variable can predict future values of another. Additionally, we use multiple time lags—half-year, one-year, two-year, and three-year—to capture potential delayed effects of one on the other. These tests are conducted for different time granularity (monthly, bi-monthly, quarterly, and bi-annual), allowing us to identify causality across different time frames and adoption scenarios.

\section{Results}
\label{result}

In this section, we present the outcomes of our analysis, including the correlations for both EV adoption and CI growth using a graph theoretic approach. We examine the two cases: \textit{Early Adoption} (Section \ref{Early}) and \textit{Late Adoption} (Section \ref{Late}) using different time granularity. Additionally, we present the results of our analysis of causality relationships between EV adoption and CI growth in these two cases in (Section \ref{causalTets}).

\begin{figure*}[htbp] 
    \centering

    \setlength{\abovecaptionskip}{0pt} 
    \setlength{\belowcaptionskip}{0pt} 

    \begin{minipage}[t]{0.48\textwidth} 
        \centering
        \fbox{\includegraphics[width=\linewidth, height=8cm]{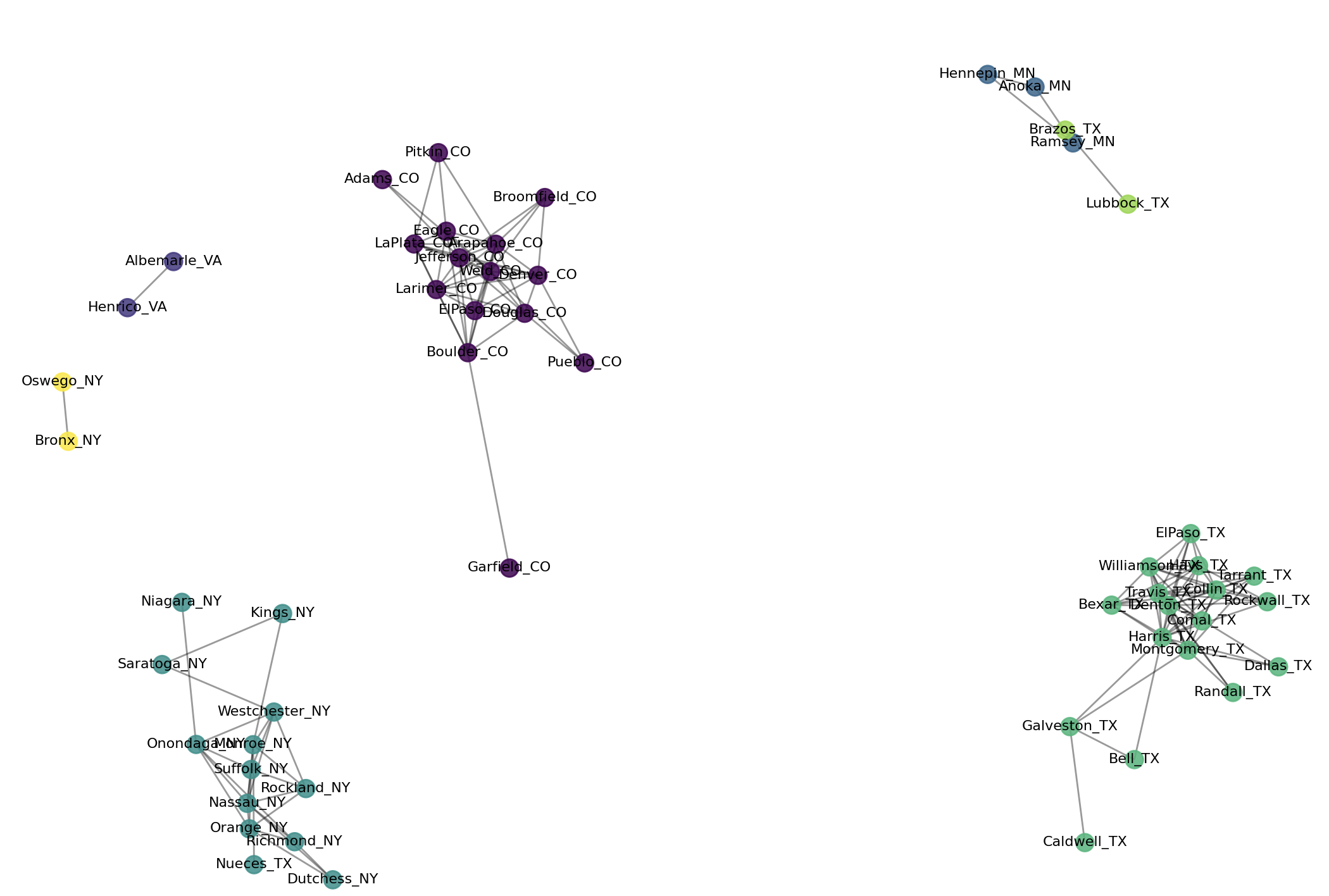}} 
        \caption*{A: Correlation Network for Monthly Data, Threshold: 0.7005}
        \label{fig:image5}
    \end{minipage}
    \hfill
    \begin{minipage}[t]{0.48\textwidth}
        \centering
        \fbox{\includegraphics[width=\linewidth, height=8cm]{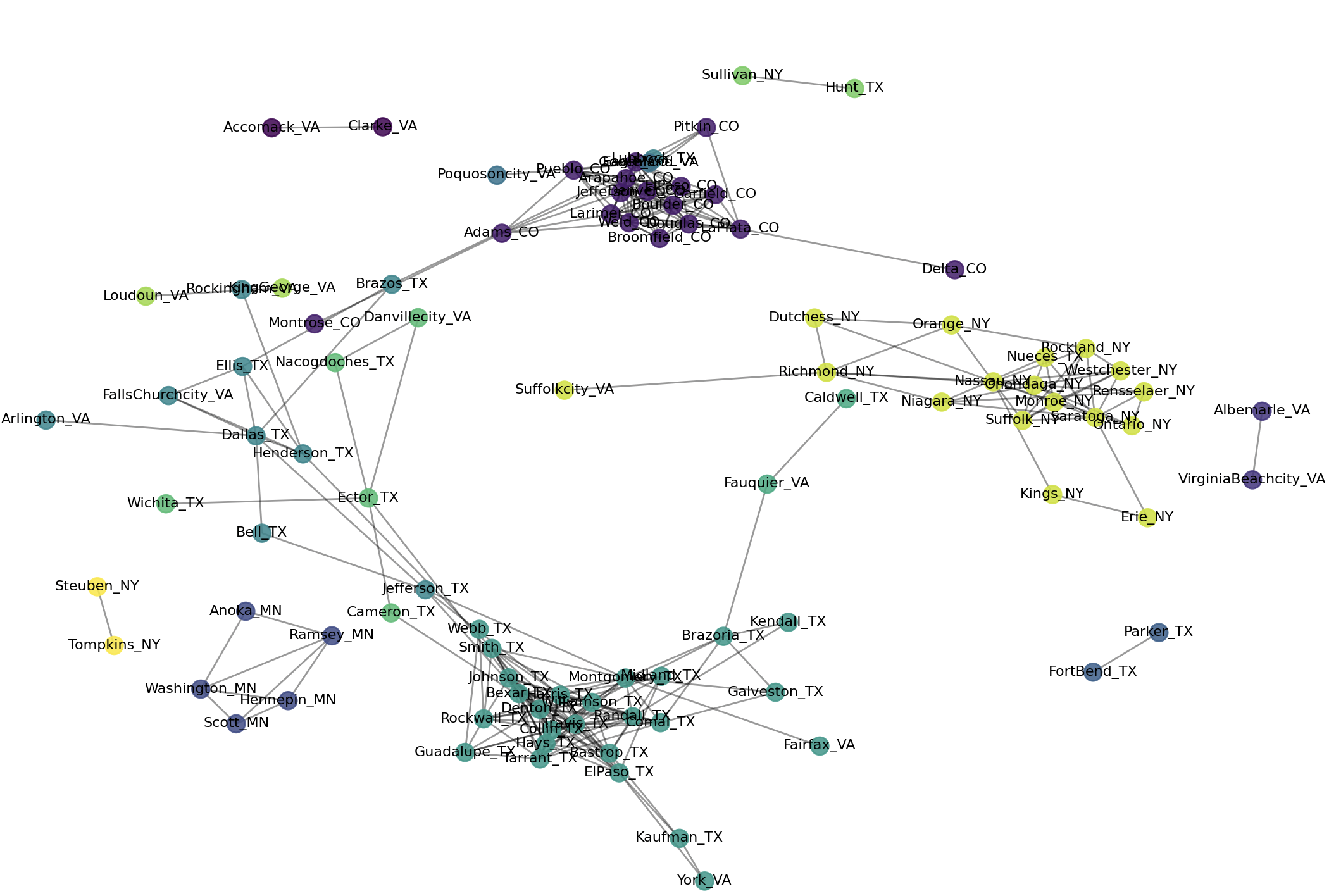}} 
        \caption*{B: Correlation Network for Bi-monthly Data, Threshold: 0.8015}
        \label{fig:image6}
    \end{minipage}

    \vspace{1cm} 

    \begin{minipage}[t]{0.48\textwidth}
        \centering
        \fbox{\includegraphics[width=\linewidth, height=8cm]{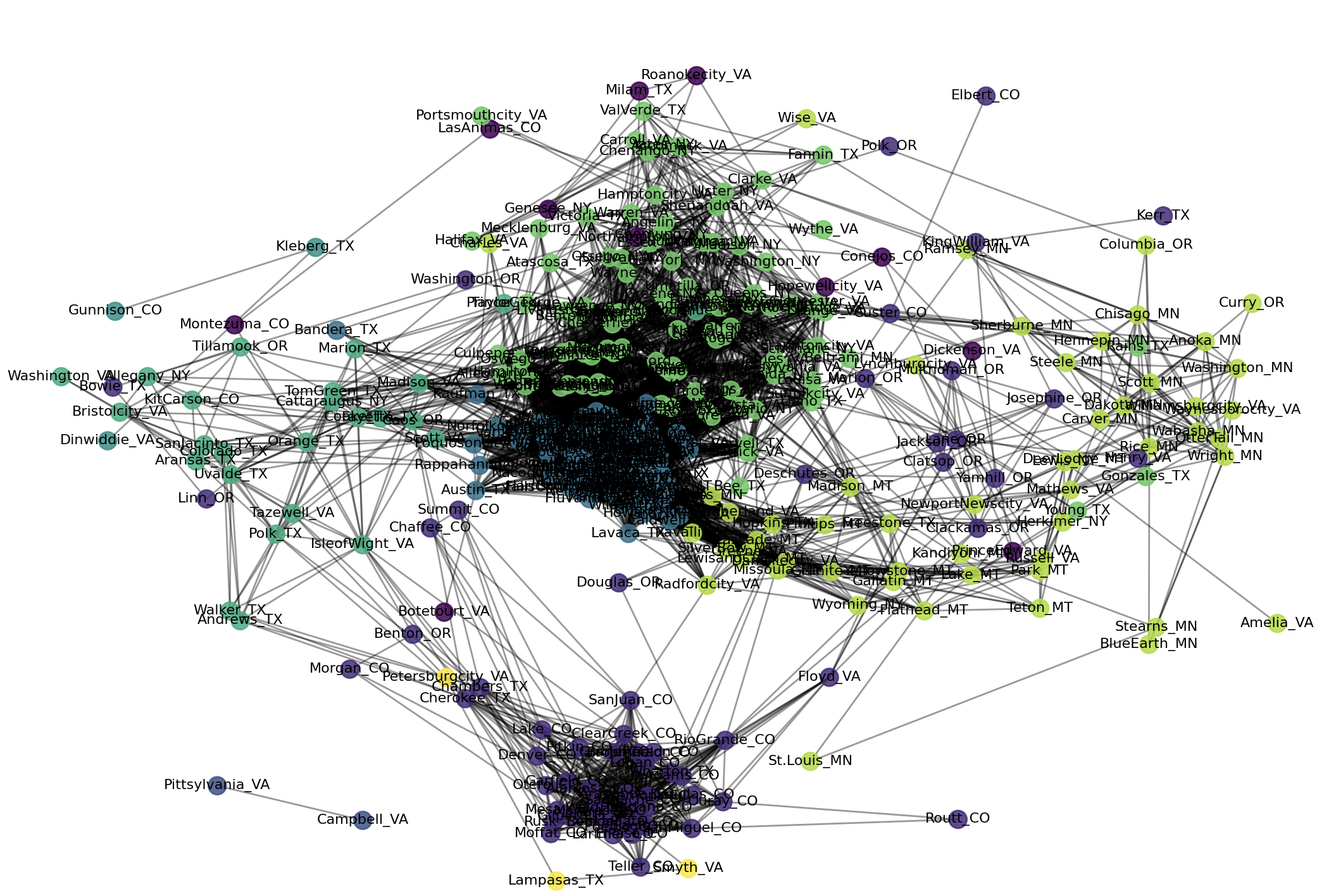}} 
        \caption*{C: Correlation Network for Quarterly Data, Threshold: 0.8379}
        \label{fig:image7}
    \end{minipage}
    \hfill
    \begin{minipage}[t]{0.48\textwidth}
        \centering
        \fbox{\includegraphics[width=\linewidth, height=8cm]{EV18ba.png}} 
        \caption*{D: Correlation Network for Bi-annually Data, Threshold: 0.8394}
        \label{fig:image8}
    \end{minipage}

    \caption{Correlation Networks of EV Adoption (2018-2022) used in \textit{Early Adoption} Scenario Using Different Time Granularities}
    \label{fig:combined1} 
\end{figure*}



\begin{figure*}[htbp]
    \centering

    \setlength{\abovecaptionskip}{0pt} 
    \setlength{\belowcaptionskip}{0pt} 

    \begin{minipage}[t]{0.48\textwidth} 
        \centering
        \fbox{\includegraphics[width=\linewidth, height=8cm]{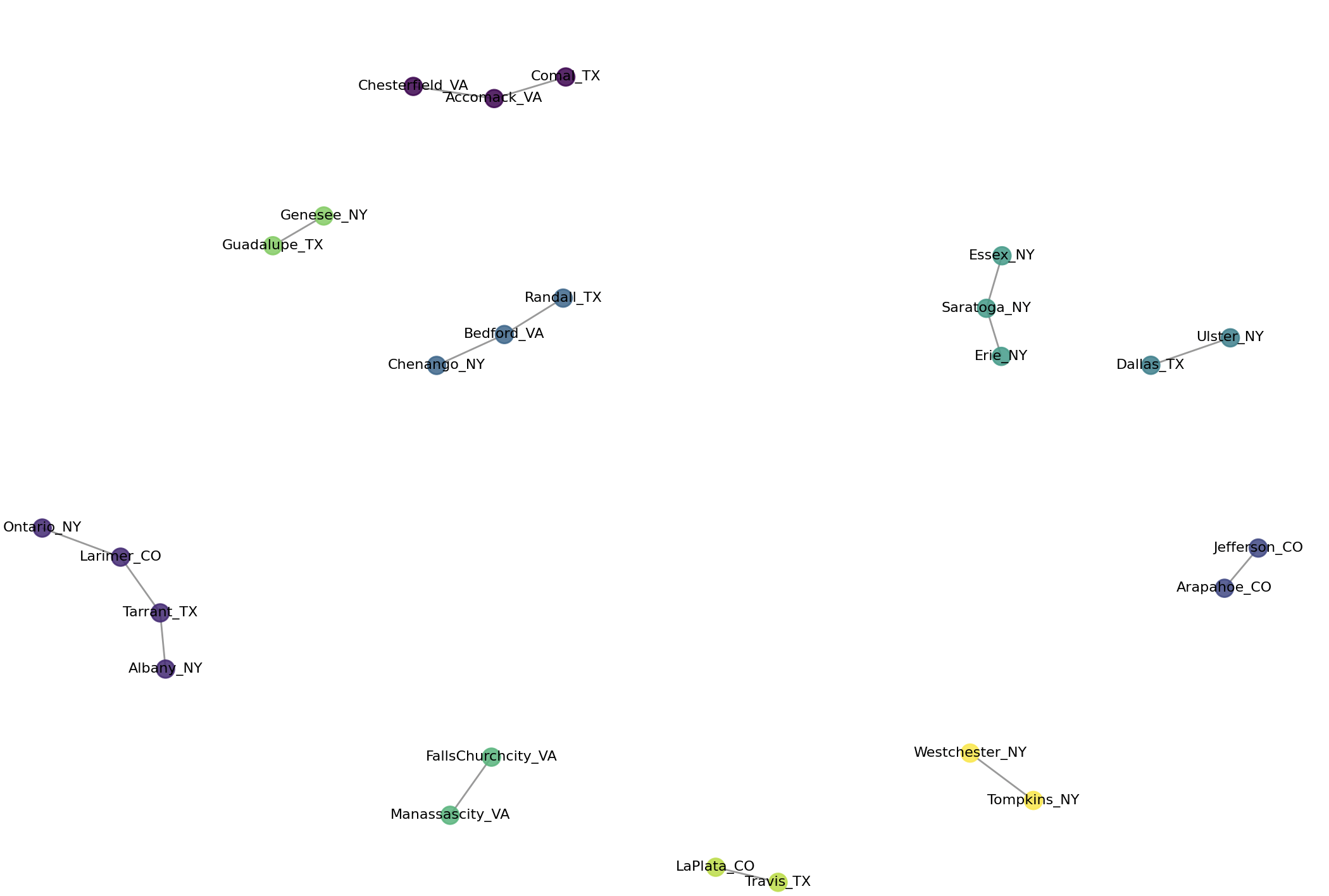}} 
        \caption*{A: Correlation Network for Monthly Data, Threshold: 0.7000}
        \label{fig:image5}
    \end{minipage}
    \hfill
    \begin{minipage}[t]{0.48\textwidth}
        \centering
        \fbox{\includegraphics[width=\linewidth, height=8cm]{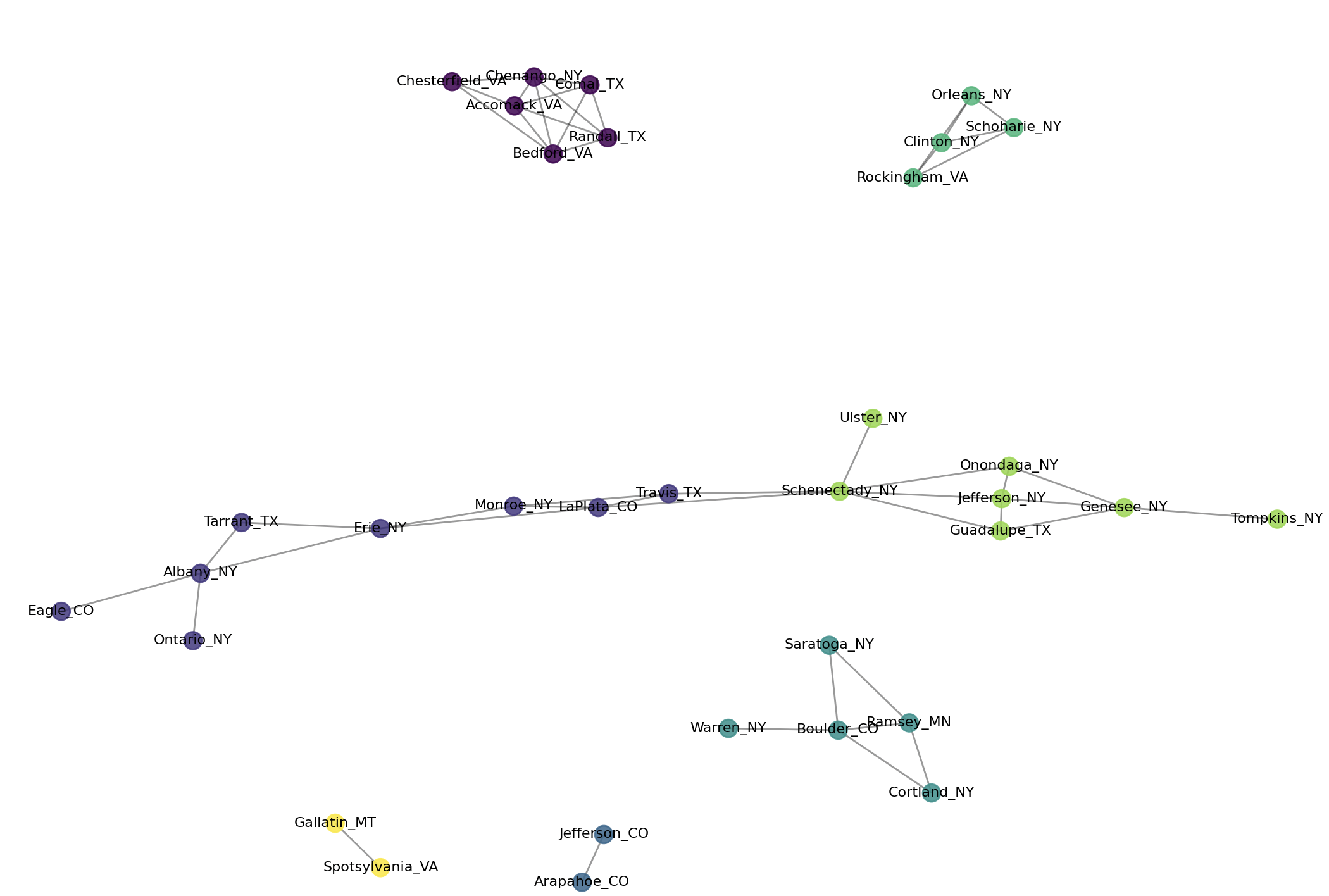}} 
        \caption*{B: Correlation Network for Bi-monthly Data, Threshold: 0.7848}
        \label{fig:image6}
    \end{minipage}

    \vspace{1cm} 

    \begin{minipage}[t]{0.48\textwidth}
        \centering
        \fbox{\includegraphics[width=\linewidth, height=8cm]{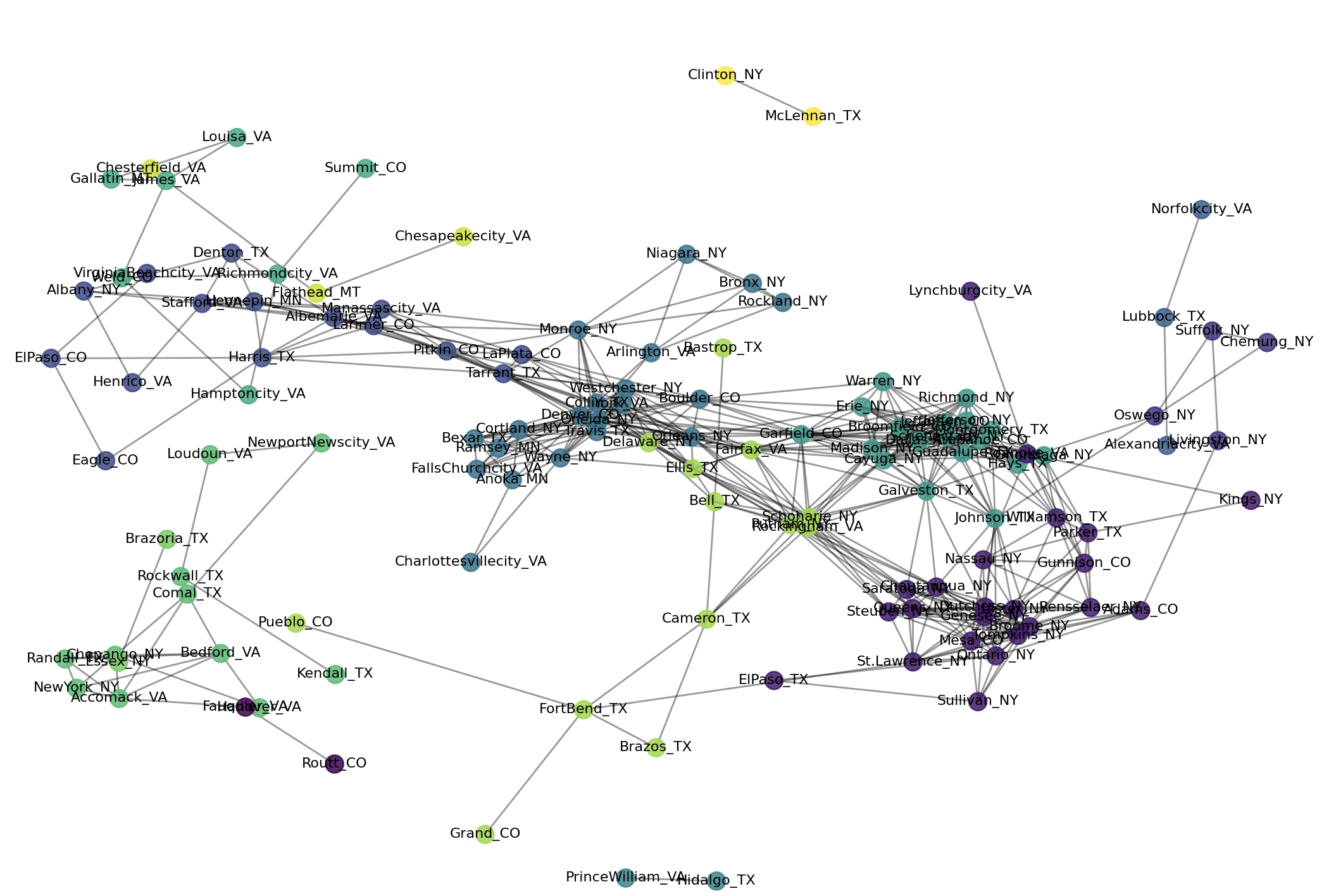}} 
        \caption*{C: Correlation Network for Quarterly Data, Threshold: 0.8439}
        \label{fig:image7}
    \end{minipage}
    \hfill
    \begin{minipage}[t]{0.48\textwidth}
        \centering
        \fbox{\includegraphics[width=\linewidth, height=8cm]{All19ba.png}} 
        \caption*{D: Correlation Network for Bi-annually Data, Threshold: 0.8500}
        \label{fig:image8}
    \end{minipage}

    \caption{Correlation Networks of ALL type of Chargers growth (2019-2023) used in \textit{Early Adoption} Scenario Using Different Time Granularities}
    \label{fig:combined4} 
\end{figure*}


\subsection{Early Adoption of EV}
\label{Early}

Graph \ref{fig:combined1} illustrates the correlation networks between counties for their EV adoptions (2018-2022) using different time granularity in the scenario of \textit{Early Adoption}. The first notable finding is that lower levels of granularity produce more homogeneous clusters. For example, on a monthly basis, we observe that each cluster comprises counties from the same state. However, as the seasonal span increases, this homogeneity decreases, though it remains noticeable. Moreover, counties within the same cluster and state tend to be geographically close to each other. This raises the question of whether geographical proximity influences EV adoption—whether areas that are close to one another are more likely to adopt EVs if one area has already adopted them extensively (Contagion of EV Adoption.)

Similarly, Graph \ref{fig:combined4} illustrates the correlation between counties regarding the “ALL” level of CI (2019-2023) in the scenario of \textit{Early Adoption} Scenario. Due to space limitations, and since the findings for “ALL” level chargers apply to other levels of CI, we do not show the other levels. In general, the clusters identified in CI do not exhibit the same patterns seen in EV adoption. The CI clusters are far more heterogeneous in terms of the states their counties belong to. The discrepancies between the EV and CI correlation patterns highlight the complexity of the relationship between EV adoption and CI development.

Lastly, Table \ref{table: table0} shows the correlation between EV adoption and various CI levels (Level 2 chargers, DC chargers, and ALL) in the scenario of \textit{Early adoption} . We observe that the correlations are consistently weakly positive. Moreover, lower levels of granularity tend to have slightly stronger correlations than higher levels.

\begin{table}[]
\centering
\caption{Correlation between the EV adoption correlation and the correlation of various charging infrastructure types (DC chargers, Level 2 chargers, and ALL chargers) for \textit{Early Adoption} Scenario, across different levels of temporal granularity.}
\begin{tabularx}{\columnwidth}{|X|l|l|l|}
\hline
Granularity  & EV vs DC & EV vs Level 2 & EV vs ALL \\ \hline
Monthly & 0.1772 & 0.2205 & 0.2305 \\ \hline
Bi-monthly & 0.0972 & 0.1345 & 0.1416 \\ \hline
Quarterly & 0.0715 & 0.0624 & 0.0543 \\ \hline
Bi-annually & 0.0123 & 0.0366 & 0.0214 \\ \hline

\end{tabularx}
\label{table: table0}
\end{table}


\begin{figure*}[htbp] 
    \centering

    \setlength{\abovecaptionskip}{0pt} 
    \setlength{\belowcaptionskip}{0pt} 

    \begin{minipage}[t]{0.48\textwidth} 
        \centering
        \fbox{\includegraphics[width=\linewidth, height=8cm]{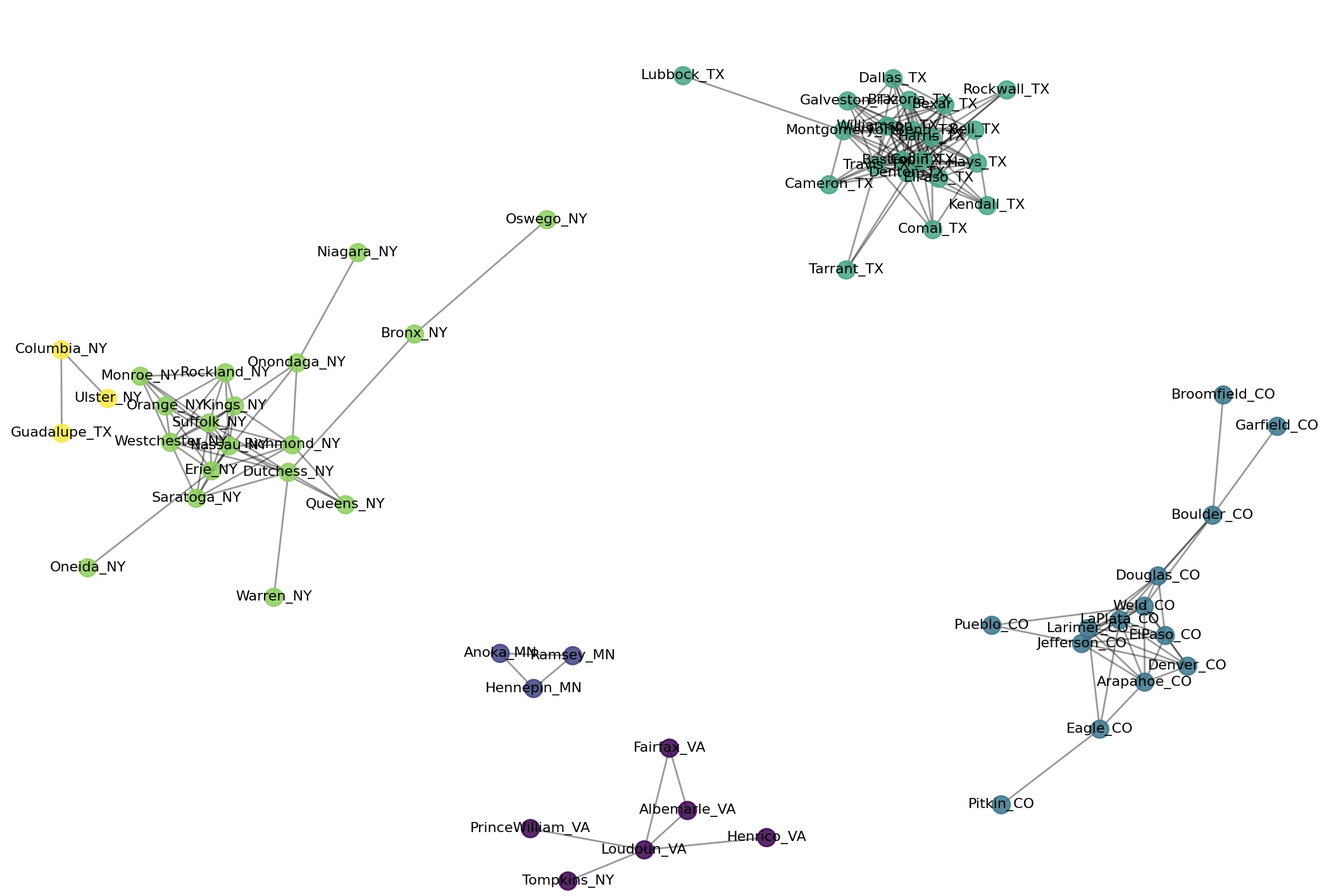}} 
        \caption*{A: Correlation Network for Monthly Data, Threshold: 0.7020}
        \label{fig:image5}
    \end{minipage}
    \hfill
    \begin{minipage}[t]{0.48\textwidth}
        \centering
        \fbox{\includegraphics[width=\linewidth, height=8cm]{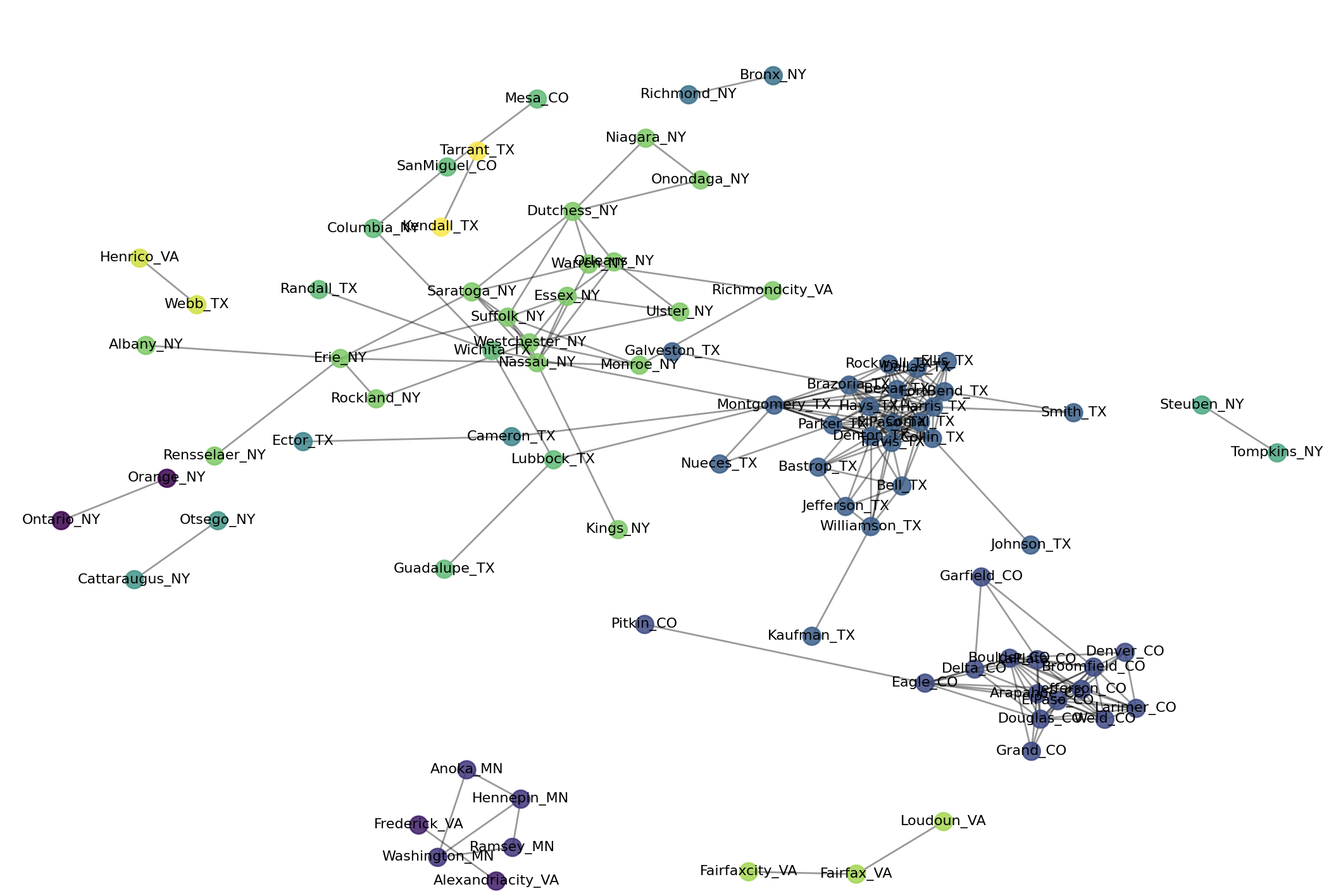}} 
        \caption*{B: Correlation Network for Bi-monthly Data, Threshold: 0.8015}
        \label{fig:image6}
    \end{minipage}

    \vspace{1cm} 

    \begin{minipage}[t]{0.48\textwidth}
        \centering
        \fbox{\includegraphics[width=\linewidth, height=8cm]{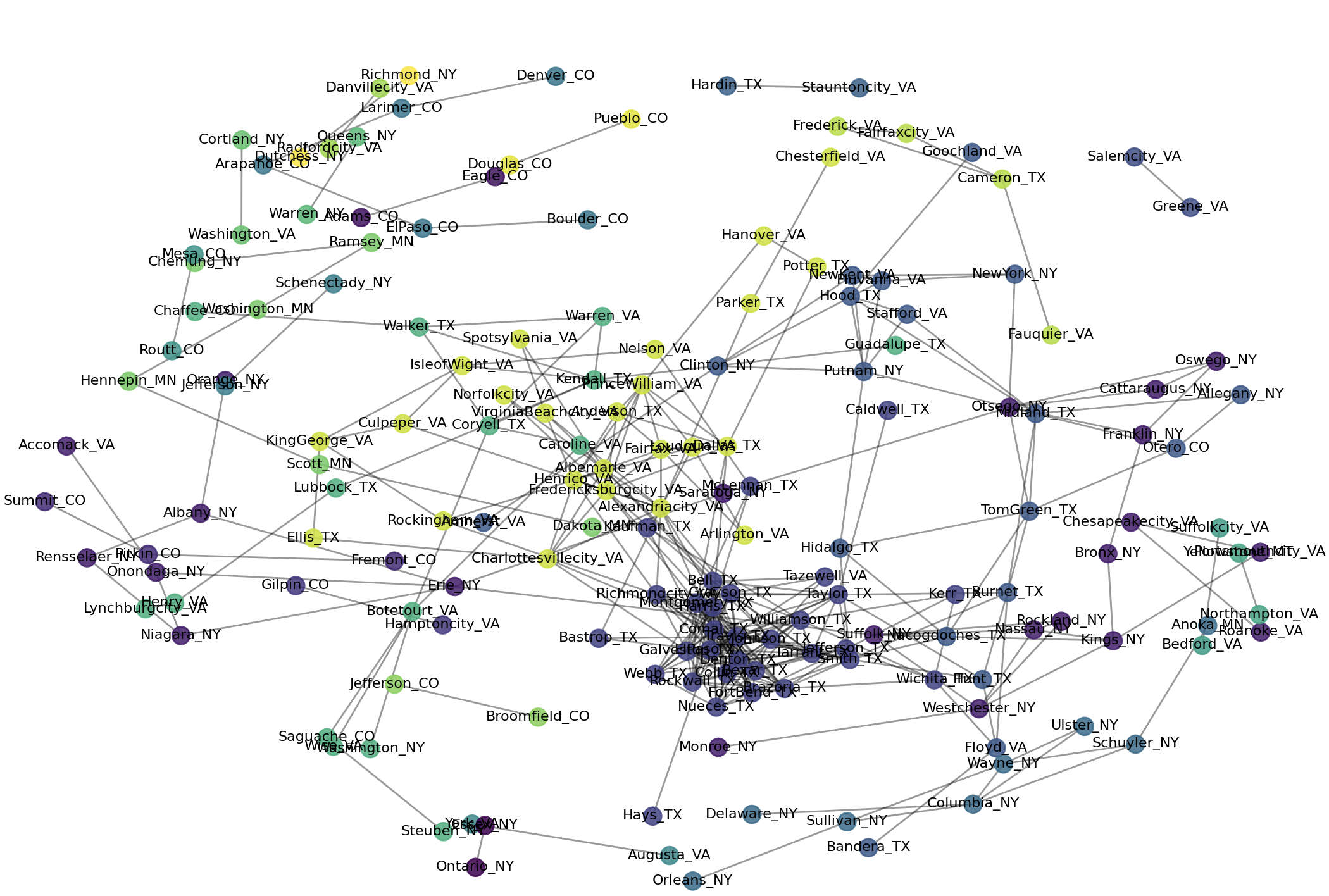}} 
        \caption*{C: Correlation Network for Quarterly Data, Threshold: 0.8379}
        \label{fig:image7}
    \end{minipage}
    \hfill
    \begin{minipage}[t]{0.48\textwidth}
        \centering
        \fbox{\includegraphics[width=\linewidth, height=8cm]{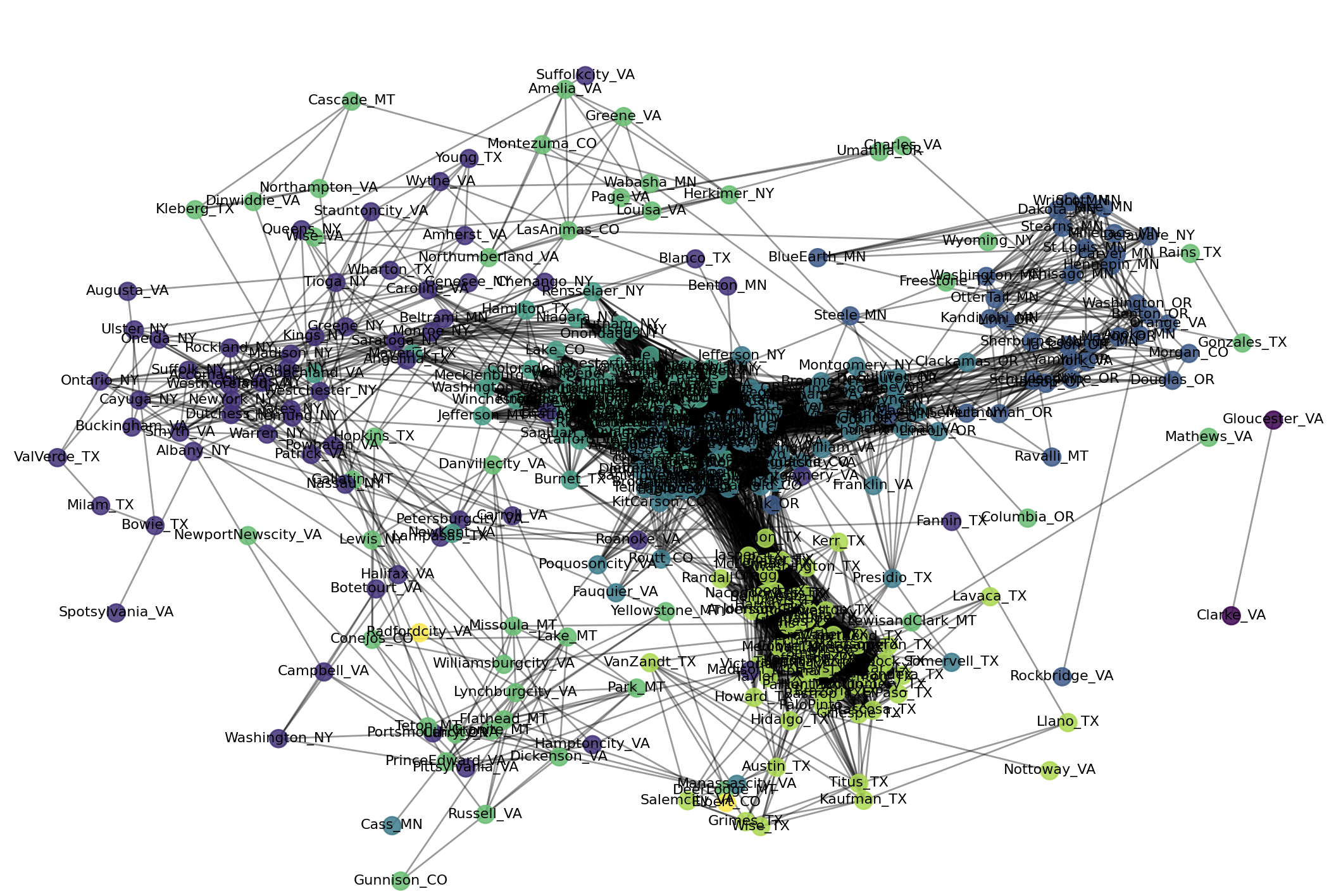}} 
        \caption*{D: Correlation Network for Bi-annually Data, Threshold: 0.8394}
        \label{fig:image8}
    \end{minipage}

    \caption{Correlation Networks of EV Adoption (2019-2023) used in \textit{Late Adoption} Scenario  Using Different Time Granularities}
    \label{fig:combined2} 
\end{figure*}



\begin{figure*}[htbp]
    \centering

    \setlength{\abovecaptionskip}{0pt} 
    \setlength{\belowcaptionskip}{0pt} 

    \begin{minipage}[t]{0.48\textwidth} 
        \centering
        \fbox{\includegraphics[width=\linewidth, height=8cm]{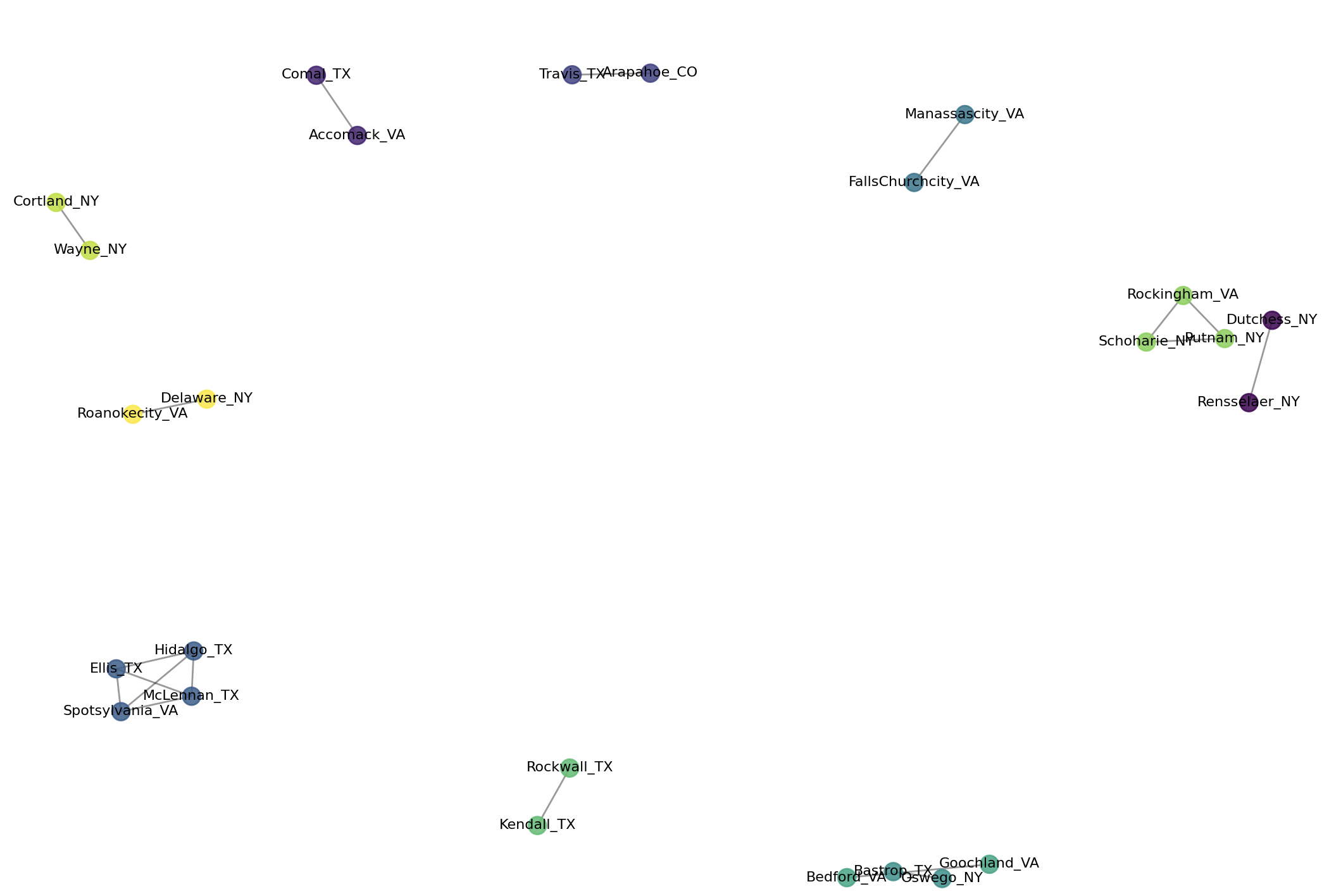}} 
        \caption*{A: Correlation Network for Monthly Data, Threshold: 0.7110}
        \label{fig:image5}
    \end{minipage}
    \hfill
    \begin{minipage}[t]{0.48\textwidth}
        \centering
        \fbox{\includegraphics[width=\linewidth, height=8cm]{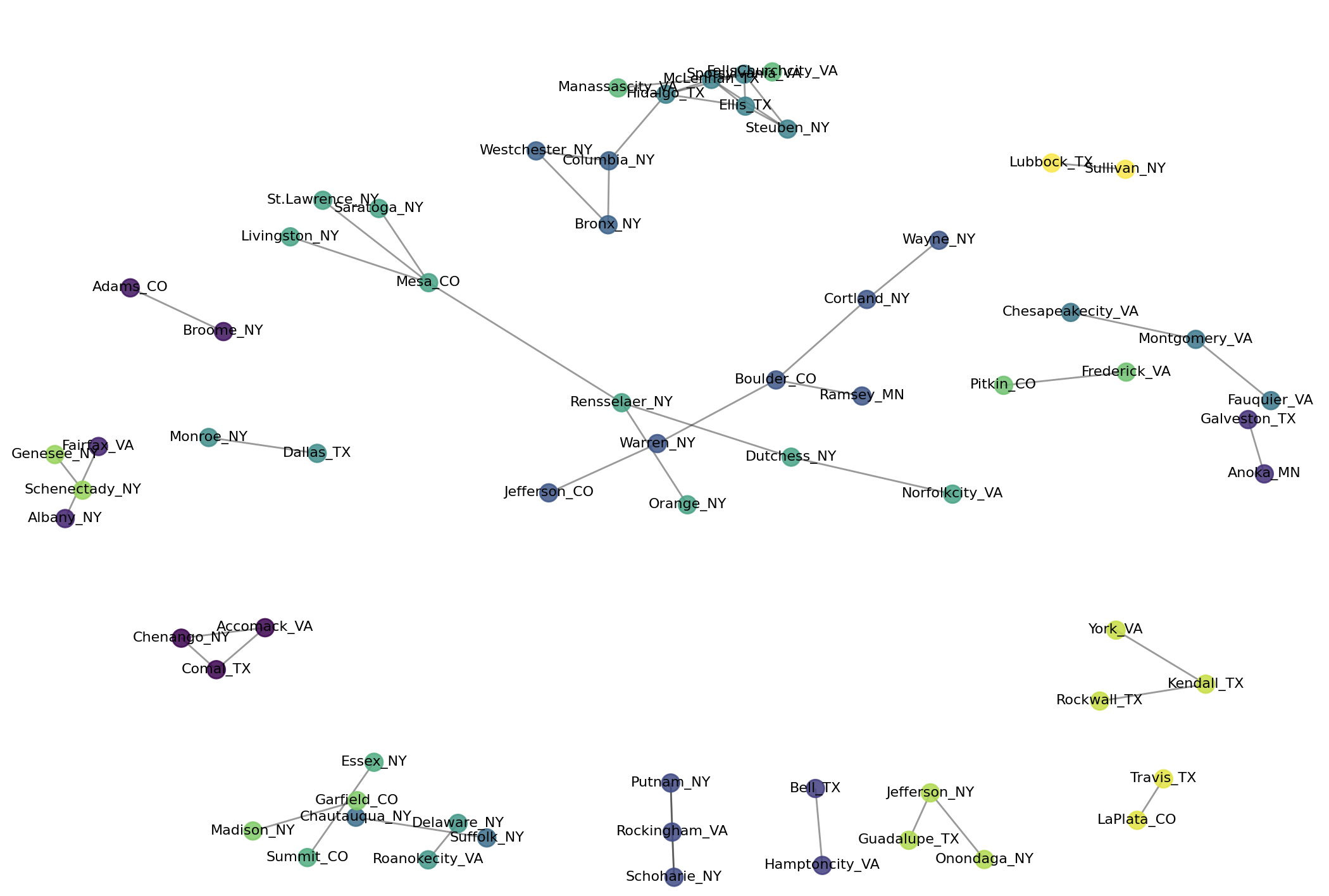}} 
        \caption*{B: Correlation Network for Bi-monthly Data, Threshold: 0.7848}
        \label{fig:image6}
    \end{minipage}

    \vspace{1cm} 

    \begin{minipage}[t]{0.48\textwidth}
        \centering
        \fbox{\includegraphics[width=\linewidth, height=8cm]{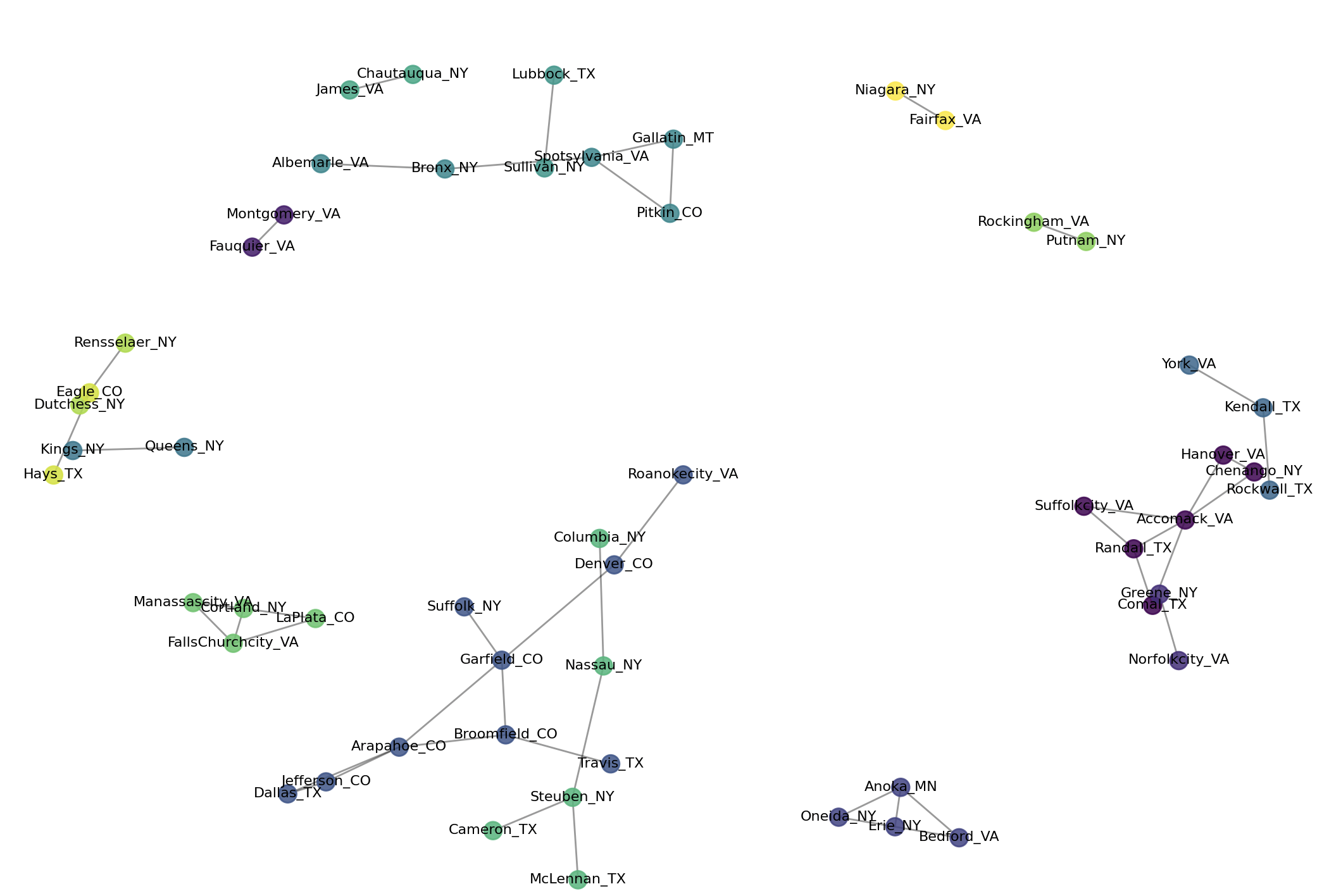}} 
        \caption*{C: Correlation Network for Quarterly Data, Threshold: 0.8439}
        \label{fig:image7}
    \end{minipage}
    \hfill
    \begin{minipage}[t]{0.48\textwidth}
        \centering
        \fbox{\includegraphics[width=\linewidth, height=8cm]{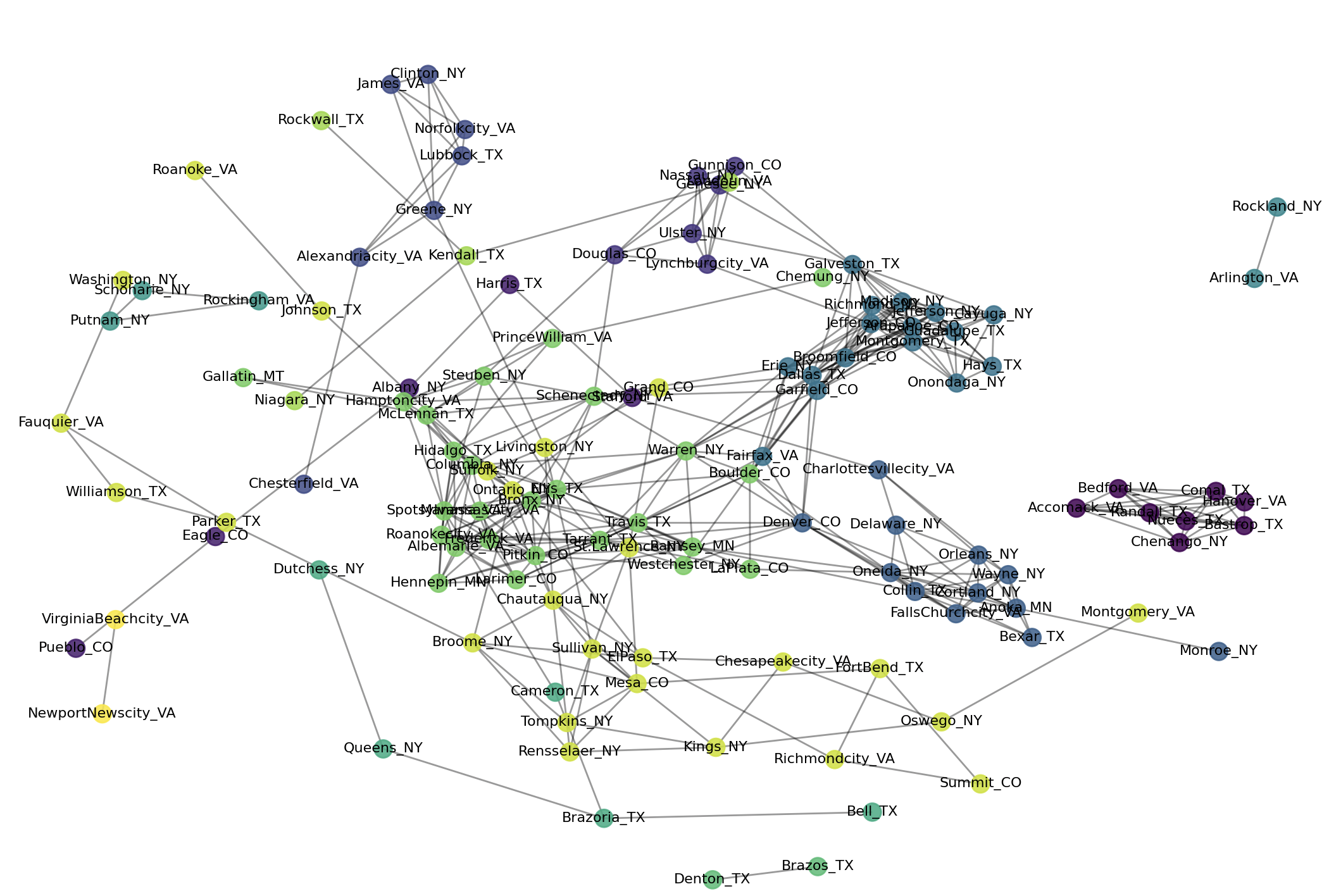}} 
        \caption*{D: Correlation Network for Bi-annually Data, Threshold: 0.8500}
        \label{fig:image8}
    \end{minipage}

    \caption{Correlation Networks of ALL type of Chargers growth (2018-2022) used in \textit{Late Adoption} Scenario Using Different Time Granularities}
    \label{fig:combined3} 
\end{figure*}

\subsection{Late Adoption of EV}
\label{Late}

Graph \ref{fig:combined2} illustrates the scenario of \textit{Late Adoption} for the EV adoption (2019-2023). A similar pattern to \textit{Early Adoption} scenario is observed, where lower levels of granularity show more homogeneity. However, there are noticeable differences between the clusters in the scenarios of \textit{Early} and \textit{Late Adoption}. One key difference is that the \textit{Early Adoption} networks tend to be slightly denser than the \textit{Late Adoption} networks. This could be due to the sparser nature of the early adoption data (2018-2022), which results in a higher number of correlations between counties.

For the correlations between counties regrading CI growth in the \textit{Late Adoption} scenario, we only show the correlation for the “ALL” level of CI for simplicity in Graph \ref{fig:combined3}. Analogously, the findings regarding CI promotion in the \textit{Late Adoption} scenario are similar to those in \textit{Early Adoption}, where the EV and CI correlation patterns differ significantly.

Lastly, Table \ref{table: table1} shows the correlation between EV adoption and various CI levels (Level 2 chargers, DC chargers, and ALL) in the \textit{Late Adoption} scenario . We observe that the correlations are consistently weakly positive. Moreover, lower levels of time granularity tend to show slightly stronger correlations. Additionally, the correlation between EV adoption and CI correlation in the \textit{Late Adoption} scenario is consistently higher than in the \textit{Early Adoption} scenario across all levels of time granularity. This supports the conclusion that providing adequate charging infrastructure is crucial to promoting the use of EVs.


\begin{table}[]
\centering
\caption{Correlation between the EV adoption correlation and the correlation of various charging infrastructure types (DC chargers, Level 2 chargers, and ALL chargers) for \textit{late adoption}, across different levels of temporal granularity.}
\begin{tabularx}{\columnwidth}{|X|l|l|l|}
\hline
Granularity  & EV vs DC & EV vs Level 2 & EV vs ALL \\ \hline
Monthly & 0.1415 & 0.2529 & 0.2667 \\ \hline
Bi-monthly & 0.1806 & 0.1347 & 0.1824 \\ \hline
Quarterly & 0.0897 & 0.0737 & 0.0619 \\ \hline
Bi-annually & 0.0263 & 0.0368 & 0.0203 \\ \hline

\end{tabularx}
\label{table: table1}
\end{table}

\subsection{Causal Relationships Between EV Adoption and Charging Infrastructure}
\label{causalTets}

\begin{table}[h!]
\centering
\caption{The number of Counties Rejecting the Null hypothesis of the Causal Relationships in the \textit{Early Adoption} Where the "EV" Causes the "ALL".}
\renewcommand{\arraystretch}{1.5} 
\setlength{\tabcolsep}{8pt} 
\begin{tabular}{|l|c|c|c|c|c|}
\hline
\textbf{Granularity\textbackslash Lag} & \textbf{half-year} & \textbf{one-year} & \textbf{two-year} & \textbf{three-year} & \textbf{Unique} \\ \hline
\textbf{monthly}    & 12 & 20 & 16 & 13 & 49 \\ \hline
\textbf{Bi-monthly} & 12 & 16 & 18 & 14 & 52 \\ \hline
\textbf{quarterly}  & 17 & 19 & 17 & 15 & 57 \\ \hline
\textbf{Bi-annual}  & 9  & 12 & 6  & 15 & 30 \\ \hline
\textbf{Unique}     & 28  & 38 & 45  & 33 & \textbf{93} \\ \hline
\end{tabular}
\label{tab:ev->all}
\end{table}

\begin{table}[h!]
\centering
\caption{The number of Counties Rejecting the Null hypothesis of the Causal Relationships in the \textit{Late Adoption} Where the "ALL" Causes the "EV".}
\renewcommand{\arraystretch}{1.5} 
\setlength{\tabcolsep}{8pt} 
\begin{tabular}{|l|c|c|c|c|c|}
\hline
\textbf{Granularity\textbackslash Lag} & \textbf{half-year} & \textbf{one-year} & \textbf{two-year} & \textbf{three-year} & \textbf{Unique} \\ \hline
\textbf{monthly}    & 18 & 11 & 9 & 9 & 37 \\ \hline
\textbf{Bi-monthly} & 23 & 9 & 4 & 12 & 39 \\ \hline
\textbf{quarterly}  & 15 & 6 & 4 & 8 & 27 \\ \hline
\textbf{Bi-annual}  & 17  & 8 & 7  & 8 & 35 \\ \hline
\textbf{Unique}     & 44  & 26 & 20  & 26 & \textbf{77} \\ \hline
\end{tabular}
\label{tab:all->ev}
\end{table}

\begin{table}[h!]
\centering
\caption{The number of Counties Rejecting the Null hypothesis of the Causal Relationships in the \textit{Late Adoption} Where the "ALL" Causes the "EV".}
\renewcommand{\arraystretch}{1.5} 
\setlength{\tabcolsep}{8pt} 
\begin{tabular}{|l|c|c|c|c|c|}
\hline
\textbf{State} & \textbf{EV->ALL} & \textbf{ALL->EV} & \textbf{Unique}  \\ \hline
\textbf{New York }    & 36 & 26 & 40  \\ \hline
\textbf{Texas} & 15 & 23 & 26  \\ \hline
\textbf{Virginia}  & 23 & 16 & 27  \\ \hline
\textbf{Colorado}  & 16  & 8 & 18   \\ \hline
\textbf{Minnesota}     & 2  & 2 & 2   \\ \hline
\textbf{Montana}     & 1  & 2 & 2   \\ \hline
\textbf{Unique}     & \textbf{93}  & \textbf{77} & \textbf{115}   \\ \hline
\end{tabular}
\label{tab:final}
\end{table}

We conducted a Granger causality test to uncover the causal relationships between EV adoption and CI growth. In this analysis, we explore whether EV adoption causes CI growth or vice versa, as in the \textit{Early Adoption} and \textit{Late Adoption} scenarios, respectively. For simplicity, we focus on one level of CI, namely "ALL," which represents all types of chargers combined. Understanding the effects of specific levels, particularly DC chargers, on EV adoption is important and will be addressed in future work. In our causality analysis, we use different time granularities, including monthly, bi-monthly, quarterly, and bi-annual periods. Moreover, we test various lags between EV adoption and CI growth, including half-year, one-year, two-year, and three-year lags, to uncover potential relationships. Hence, each combination of a time granularity, a time lag, and an adoption scenario is tested for each county in our analysis. The number of counties rejecting the null hypotheses for \textit{Early Adoption} and \textit{Late Adoption} as explained in Section \ref{causalTets}, is detailed in this section. 

Table \ref{tab:ev->all} presents the causal relationships captured in the \textit{Early Adoption} scenario, where EV adoption causes CI "ALL" growth. Each entry in Table \ref{tab:ev->all} represents a combination of a time granularity and time lag, indicating the number of counties with significant evidence of causal relationships. For instance, when using quarterly granularity and a two-year lag, 17 counties show such causal relationships. To account for counties appearing repeatedly across different combinations, the "Unique" column represents the unique counties showing causal relationships for a specific time granularity across different time lags, while the "Unique" row indicates the unique counties for a time lag across different granularities. Similarly, Table \ref{tab:all->ev} presents the causal relationships captured in the \textit{Late Adoption} scenario, where CI "ALL" growth causes EV adoption.

From Tables \ref{tab:ev->all} and \ref{tab:all->ev}, we observe that the number of unique counties in the \textit{Early Adoption} scenario (\textbf{93 out of 137}) is larger than in the \textit{Late Adoption} scenario (\textbf{77 out of 137}). This indicates that EV adoption tends to cause CI growth more frequently than CI growth causes EV adoption. Consequently, EV adoption in an area before CI development is more critical than investing in CI before EV adoption. Thus, focusing on incentivizing EV purchases is a significant step for widespread adoption. Nevertheless, the demonstration of CI growth causing EV adoption in several counties highlights the complex and bidirectional nature of the relationship between EVs and CI.

Secondly, the causal effect of EV adoption on CI growth is clearly slower than the reverse. Specifically, the "Unique" row in Table \ref{tab:ev->all} shows that the peak number of counties (\textbf{45 counties}) exhibiting CI growth caused by EV adoption occurs at a two-year lag. In contrast, the causal effect of CI growth on EV adoption peaks at a half-year lag (\textbf{44 counties}) as shown in Table \ref{tab:all->ev}. This finding is intuitive, as the slower causal effect of EV adoption on CI growth is likely due to the high costs associated with investing in CI infrastructure, such as charging stations. On the other hand, the causal effect of CI growth on EV adoption is faster, as purchasing an EV is an individual decision and typically less costly than establishing CI.

Overall, Table \ref{tab:final} shows that 115 unique counties out of 137 (approximately 84\% of the data) exhibit significant causal relationships between EV adoption and CI growth using various time granularities, time lags, and adoption scenarios. Furthermore, Table \ref{tab:final} reveals that three states have a higher number of counties where EV adoption causes CI "ALL" growth, two states have more counties where CI "ALL" growth causes EV adoption, and one state exhibits an equal number of counties for both causal relationships.

\subsection{Outlier Counties in EV Adoption Clusters}

Focusing on the reasons behind the appearance of counties from different states within a cluster dominated by another state requires a more in-depth extension of this research. For instance, in the early adoption scenario using bi-monthly data, Danville City in Virginia appears in a cluster primarily composed of Texas counties. In this specific case, we identified shared factors such as rural influence, moderate commute distances, limited access to public transit, and lower median income, which may help explain similar patterns of EV adoption.

Conducting deeper analyses of these uncommon county appearances, both at the cluster level and across different adoption scenarios (\textit{Early} vs. \textit{Late Adoption}) can offer more nuanced insights into the factors driving EV adoption. Exploring these outliers may reveal underlying conditions that are not necessarily tied to geography or state policy but rather to shared structural or socioeconomic characteristics.

\section{Limitations and Future Work}

\label{limitation}
This study focuses exclusively on the correlation and causal dynamics between electric vehicle (EV) adoption and the growth of charging infrastructure (CI), using a graph-theoretic model and time-series causality analysis. One inherent limitation is that the analysis does not explicitly isolate the influence of external factors such as policy interventions, technological advancement, and socio-demographic variables. Instead, we assume that such influences are implicitly embedded in the observed trends of EV adoption and CI growth across counties. For example, government incentives or socioeconomic conditions that encourage EV adoption are presumed to be reflected in the adoption data; likewise, incentives or budget allocations for infrastructure development are assumed to be manifested in the CI growth data. Although this assumption allows us to study the direct relationship between EVs and CIs, it restricts our ability to attribute causal mechanisms to specific policy or technological factors.



While our graph-theoretic framework reveals meaningful clustering patterns in both EV adoption and CI growth, it does not directly explain why certain counties tend to group together. Understanding the underlying drivers of these clusters—such as shared socio-economic profiles, policy environments, or infrastructure investments—remains an important next step. Future work should aim to uncover these commonalities by integrating additional contextual variables (e.g., population density, median income, political orientation, or state-level incentives) into the analysis. Whether through multivariate modeling or cluster enrichment techniques, such extensions would help determine whether the structural patterns observed in the networks reflect deeper systemic factors, thereby improving both interpretability and policy relevance. Furthermore, examining the role of different CI levels (e.g., Level 2 vs. DC fast chargers) in shaping adoption patterns under early and late adoption scenarios could offer a more granular understanding of the interplay between EV demand and infrastructure supply.

\section{Conclusion and Future Directions}

\label{conclusion}

This study employs a graph-theoretic and time-series causality approach to investigate the relationship between EV adoption and charging infrastructure (CI) growth across 137 counties in six U.S. states. The results reveal consistent clustering patterns in EV adoption, particularly within states, whereas CI growth networks show weaker structural patterns. The correlation between EV and CI growth is generally weak across early and late adoption scenarios, with slightly stronger values observed when sufficient CI precedes EV adoptxion.

Our causality analysis indicates that while EV adoption does causally influence CI growth, the effect is slower compared to the reverse scenario, where CI growth more promptly triggers EV adoption. Importantly, over 80\% of the counties analyzed exhibit significant causal relationships under at least one configuration of time lag, time granularity, and adoption scenario.

Despite the innovation in methodology, our analysis assumes that external drivers such as policy, economic incentives, and technology improvements are embedded within the temporal trends of the variables studied. Future work should incorporate multivariate analysis to isolate these effects and offer more actionable policy recommendations. Additionally, extending this framework to include county-level attributes may uncover structural explanations behind the observed clustering, enhancing our understanding of spatial and temporal patterns in EV adoption and CI expansion.

\bibliographystyle{unsrt}  


\end{document}